\documentclass[prb,floatfix,twocolumn,showpacs,amsmath,amssymb,superscriptaddress,titlepage]{revtex4-1}
\usepackage{graphicx}
\usepackage{epstopdf}
\usepackage{epsfig}
\usepackage{dcolumn}
\usepackage{bm}
\usepackage{color}
\usepackage{subfigure}
\usepackage{wrapfig}
\usepackage{rotate,color,url}
\usepackage{longtable}
\usepackage{time}
\usepackage[pagebackref=false,colorlinks,linkcolor=blue,citecolor=blue,urlcolor=magenta]{hyperref}

\begin{document}
\newcommand {\be}{\begin{equation}}
\newcommand {\ee}{\end{equation}}
\newcommand {\bea}{\begin{array}}
\newcommand {\cl}{\centerline}
\newcommand {\eea}{\end{array}}
\newcommand {\pa}{\partial}
\newcommand {\al}{\alpha}
\newcommand {\de}{\delta}
\newcommand {\ta}{\tau}
\newcommand {\ga}{\gamma}
\newcommand {\ep}{\epsilon}
\newcommand {\si}{\sigma}
\newcommand{\up}{\uparrow}
\newcommand{\down}{\downarrow}

\title{Three phase classification of  an uninterrupted traffic flow: a $k$-means clustering study}

\author{Reihaneh Kouhi}
\email{r.koohiesfahani@ph.iut.ac.ir}
\affiliation{Department of Physics, Isfahan University of Technology, Isfahan 84156-83111, Iran}

\author{Farhad Shahbazi}
\email{shahbazi@cc.iut.ac.ir}
\affiliation{Department of Physics, Isfahan University of Technology, Isfahan 84156-83111, Iran}
\affiliation{School of Physics, Institute for Research in Fundamental Sciences (IPM), Tehran 19395-5531, Iran}

\author{Meisam Akbarzadeh}
\email{makbarzadeh@cc.iut.ac.ir}
\affiliation{Department of Transportation Engineering, Isfahan University of Technology, Isfahan 84156-83111, Iran}

\date{\today}

\begin{abstract}
We investigate the speed time series of the vehicles recorded  by a camera at a section of a highway in the city of Isfahan, Iran.
Using $k$-means clustering algorithm, we find that the natural number of clustering for this set of data is $3$. 
This is in agreement with the three-phase theory of uninterrupted traffic flows. According to this theory, 
the three traffic phases are categorized as {\em free flow} (F), {\em synchronized} (S) and {\em wide moving jam} (J). We obtain the transition speeds and densities at F-S and also S-J transitions. We also apply the Shannon entropy analysis on the speed time series over finite windows, which  equips us to monitor in areal time  the  instant state of a  traffic flow.

\end{abstract}

\maketitle
\section{Introduction \label{int}}
Understanding the nature of traffic congestion and the mechanism of its formation in uninterrupted facilities enlightens the traffic science and paves the way for more accurate short term forecasting and traffic management and control \cite {Raj, golob}.
Various analytical models proposed in this regard could be categorized in two paradigms. The first is based on fundamental diagrams (Lighthill and Whitham 1955 \cite{Lighthill}, Richards 1956 \cite{Richards}, Daganzo 1994\cite{Daganzo4}, 1995\cite{Daganzo5}, Tang et al. 2012 \cite{Tang}) and the second is based on driver reaction characteristics (Chandler et al. 1958 \cite{Chandler58} , Gazis et al. 1961\cite{Gazis}, Gipps 1981\cite{Gipps}, Krau$\beta$ 1998\cite{Krau}, Aw and Rascle 2000 \cite{Rascle}, Newell 2002 \cite{Newell}, Kesting and Treiber 2008\cite{Kesting} ).

The classical traffic theory distinguishes between two phases of traffic stream namely free and congested flow. Accordingly, the flow-density space i.e. fundamental diagram is assumed to consist of two sub-spaces. The quest for theories capable of better explaining the critical features such as traffic breakdown (onset of congestion in an initial free flow) yielded the development of the three phase traffic theory
 \cite {Kerner2, Kerner3, Kerner4, kerner}. Since its emergence, the three-phase theory has been noticeably applied in traffic analysis. Hausken and Rehborn (2015)\cite{Hausken} started extending the three-phase theory into the game theoretic domain. Jia et al. (2011)\cite{Jia} proposed a cellular automaton model with time gap dependent randomization under the three-phase traffic theory. Tian (2012)\cite{Tian} proposed average space gap model (ASGM) - a cellular automaton for traffic flow within the fundamental diagram - which could reproduce aspects of the three-phase theory. Some empirical examples of the validation of the traffic phase can be found in Rehnborn et al. (2011)\cite{rehborn} and Sch\"{a}fer et al. (2011)\cite{schafer}.
 
 According to Sch\"{o}nhof and Helbing (2009)\cite{Helbing}, a major shortcoming of the three-phase theory is its vagueness in defining the boundaries of congested phases i.e. S and J. Therefore, this paper tries to develop a framework for distinguishing three phases without having to consider too many criteria. 

In this work we employ a specific clustering scheme, the so called $k$-means clustering algorithm,  for  classification  of different phases of an uninterrupted traffic flow. 
Data for this research came from automatic speed detectors installed in urban highways of Isfahan, Iran.

The rest of the paper is organized as follows. In section ~\ref{TPTT} a brief introduction of the three phase traffic theory is presented. In section \ref{methods} the methodology used in our data analysis is introduced.   section~\ref{data} gives some details related to the recorded speed data. Sections \ref{k-mean} and \ref{Shannon} discuss  the results of applying $k$-means clustering method and also Shannon entropy on the speed time series, respectively. Section \ref{conclusion} is devoted to the conclusion.  

\section{The Three-Phase Traffic Theory}
\label{TPTT}

Three-phase traffic theory assumes free flow (F), synchronized flow (S) and wide moving jam (J) for traffic stream. Actually, under three-phase formalism, classical congested traffic is divided into two distinct phases named as synchronized flow and wide moving jam. F is the state in which because of the low density of traffic, the interactions among vehicles are negligible and vehicles move at their desired speed. S is the state in which speed is synchronized within and among lanes and there is a continuous flow with no significant stopping and low probability of passing which results in bunching of vehicles (a tendency towards synchronization of vehicle speeds in each of the road lanes). J is the state in which a jam moves upstream through any highway bottlenecks, maintaining the mean speed of the downstream front. The difference of a wide and a narrow moving jam lays in maintaining the mean speed of the downstream jam front \cite{Hausken}.

By establishing three traffic phases, three phase transitions may be imagined (F$\rightarrow$S, S$\rightarrow$J, F$\rightarrow$J) among which the theory rejects the possibility of F$\rightarrow$J. Therefore, free-flow state transits to wide moving jam only through synchronized state.
The phase transition F$\rightarrow$S at a bottleneck can be explained by a competition between over-acceleration and speed adaptation. The lane changing behavior exhibits dual roles in phase transitions. The lane changing behavior that causes a strong reduction in following vehicle speeds in the target lane is responsible for the phase transition F$\rightarrow$S or S$\rightarrow$J. In contrast, lane changing to a faster lane can lead to the phase transition S$\rightarrow$F or J$\rightarrow$S (Kerner and Klenov, 2009\cite{kerner}).
Based on three-phase theory, traffic breakdown (F$\rightarrow$S transition) occurs if two conditions are satisfied. First, the flow rate in free flow downstream ($q$) of a bottleneck lies between a threshold flow rate for traffic breakdown at the bottleneck ($q_{th}^{\bf B}$) and the maximum flow rate in free flow downstream of the bottleneck ($q_{\bf max}^{\bf free B}$). Second, a nucleus, required for traffic breakdown, appears in free flow at the bottleneck. A nucleus is a local disturbance in free flow within which the vehicle speed is equal  or lower than the critical speed or the vehicle density is equal to or greater than the critical density required for the traffic breakdown. This implies that there is infinite number of flow rates in free flow downstream of the bottleneck for which traffic breakdown at the bottleneck is possible.

It is possible to define an empirical minimum velocity ($V_{\rm min}^{\rm free,emp}$) which separates the free flow from the synchronized flow as 
\begin{equation}
V_{\rm min}^{\rm free,emp}=q_{\rm max}^{\rm free,emp}/D_{\rm max}^{\rm free,emp},
\end{equation}
which implies  that at the limiting  (maximum) point of free flow, the density and flow rate reach their maximum values denoted by  $D_{\rm max}^{\rm free, emp}$ and $q_{\rm max}^{\rm free, emp}$, respectively. At F state, speeds are greater and at S and J speeds are less than $V_{\rm min}^{\rm free, emp}$ \cite {Kerner2, Kerner3}. One can show all these information in the fundamental diagram but there is no specified region to separate S state from J state in this diagram. Three-phase theory gives a wide region in which transition from S$\rightarrow$J is possible.

To distinguish between S and J,  a parameter $I={\tau_J}/{\tau^{(a)}_{\rm del,jam}}$ is defined where $\tau_J$ is the duration of a wide moving jam and $\tau^{(a)}_{\rm del,jam}$ is the mean time in vehicle acceleration at the downstream front of a wide moving jam. Whenever $I\gg1$ and a nucleus appears, the S$\rightarrow$J transition  takes place. Although clear, this definition is not easy to use. Calculating parameter $\tau_J$  is not easy and also one needs continuous online observation of the highway to check the creation of nucleus.
\section{Methodology}
\label{methods}
Data analysis is a process of obtaining useful information from raw data. For this purpose, statistical methods  are the natural tools\cite{judd}. 
In this work, we use  the  $k$-means  clustering algorithm  and Shannon entropy to verify the three-phase states in an uninterrupted traffic flow.

\subsection{$k$-means  clustering }
We use $k$-means  clusterings  method to determine the different states of traffic flow.
$k$-means algorithm is widely used   in data mining for partitioning of $n$ measured quantities  into $k$ clusters~\cite{James,Hartigan}. 

In this method, a given  set of  measurements $(x_1, x_2,\dots, x_n)$, can be grouped in $k$ clusters by minimizing the within-cluster sum of squares (WCSS), i.e $\min_S \sum_{i=1}^k\sum_{x \in S_i}||x-\mu_i||^2$,
in which $\mu_i$ is the mean of variables  in the cluster $S_i$.

The quality and naturalness  of the clustering in $k$-means method  can be evaluated by  {\em Silhouettes} analysis~\cite{Hennig,silhouettes}. 
Silhouettes coefficient is defined as 
\begin{equation}
S(i)=\frac{b(i)-a(i)}{\max{(b(i),a(i))}},
\end{equation}
where $a(i)$ is the average distance from one data to all other data within the same cluster and $b(i)$ is  the average distance of  the data to the nearest cluster of its own assigned  cluster. 
$S_i$ varies from $-1$ to $+1$, where a high positive (negative) value indicates that the data is a well (poorly) clustered. In the case of highly negative $S_i$ one needs  to move the data to the neighboring cluster.  When $S(i)$ is close to zero,  object i can be considered to be in both clusters.  

After finding $S(i)$ for each node, one can find the average Silhouettes coefficient for all the data. The best number of clusters, known as  {\em natural} cluster numbering of that data, is the one with the biggest average Silhouettes coefficient~ \cite{Hennig}.

\subsection{Shannon Entropy}
To distinguish the different traffic states, we also use the Shannon entropy as a measure of information content of the speed time series.   
Entropy has important physical implications as the amount of "disorder" in a system \cite{shannon}. 
 Shannon Entropy is used in different branches of science such as  the structure of ecosystems \cite{Masahiko}, security valuation in finance \cite{Stutzer, Lillo} and also in Seismology \cite{Beenamol} . 

The Shannon entropy corresponding to the stochastic variable  is defined as 
\begin{equation}\label{se}
H=-\sum_i p(x_i)\log_2 (p(x_i)),
\end{equation}
 where $p(x_i)$ is the probability that $X$ takes the value $x_i$. 

In an ordered state for which $p(x_n)=1$ and $p(x_i)=0$ for any $i\neq n$ the Shanon entropy vanishes.  
 The entropy is maximum if all the states have the same probability which is equivalent to the maximum information contained in the system.

\section{Data Set}
\label{data}
Our data set is collected from a section of Sayyad highway in Isfahan, Iran. 
Our database  contains the  speed of vehicles (in the unit of  km/h), collected in $5$ days: $8-10-2014$,  $9-10-2014$ (weekend), $10-10-2015$ (weekend), $11-10-2015$ and $12-10-2015$. 
Recording limits are $5$ and $119$ km/h with the precision of $1$ km/h. 

\begin{figure}
  \centering
  \includegraphics[width=0.71\columnwidth]{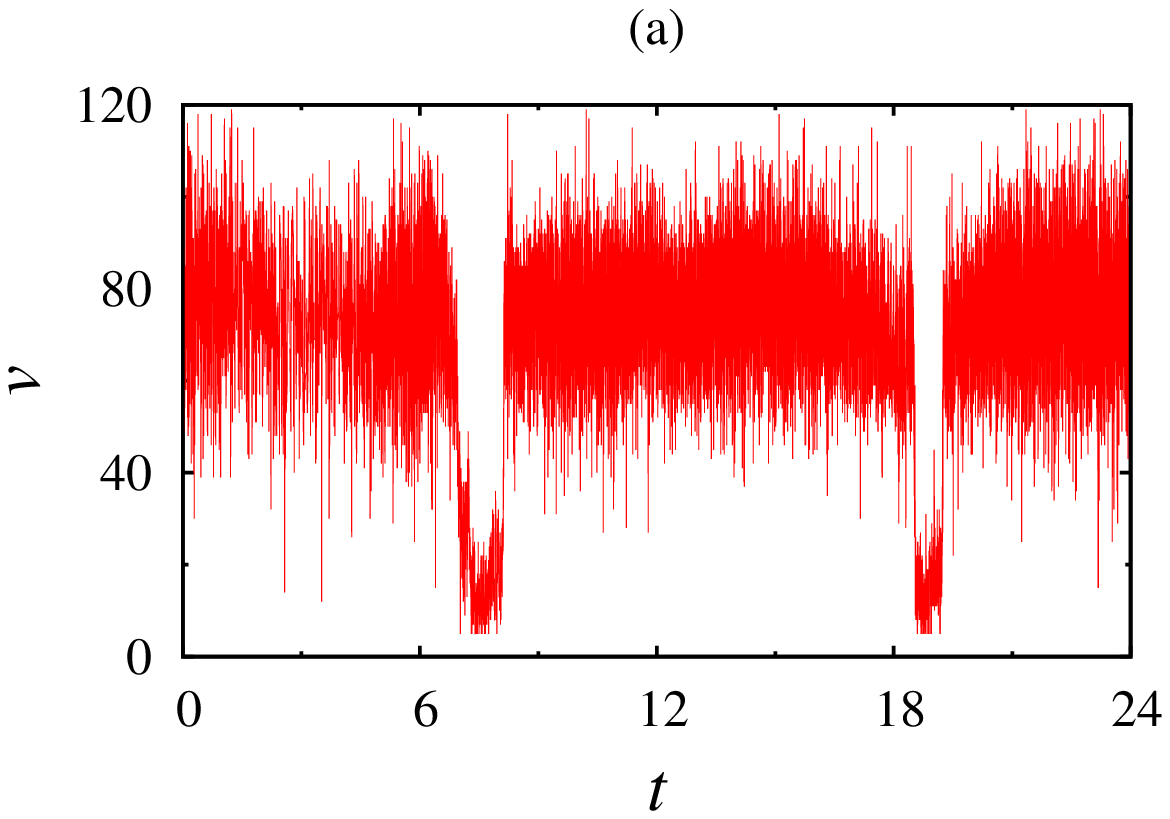}
  \includegraphics[width=0.71\columnwidth]{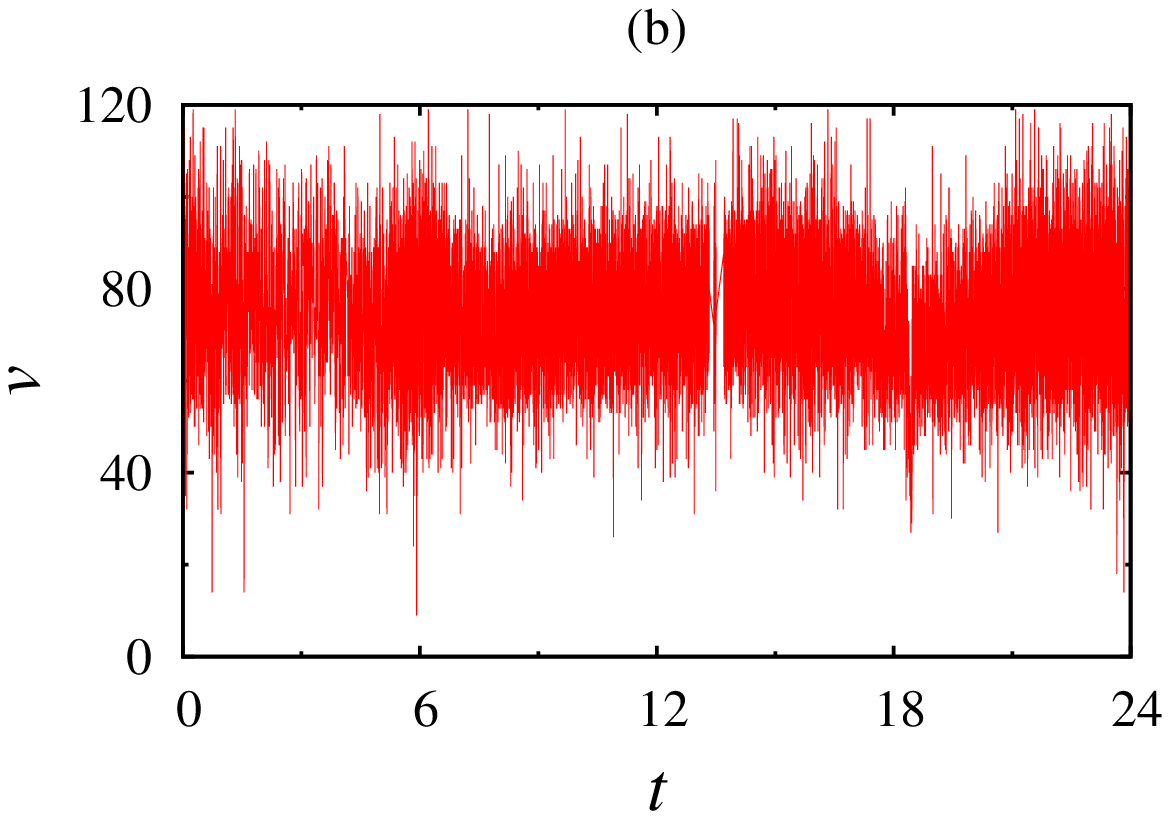}
  \includegraphics[width=0.71\columnwidth]{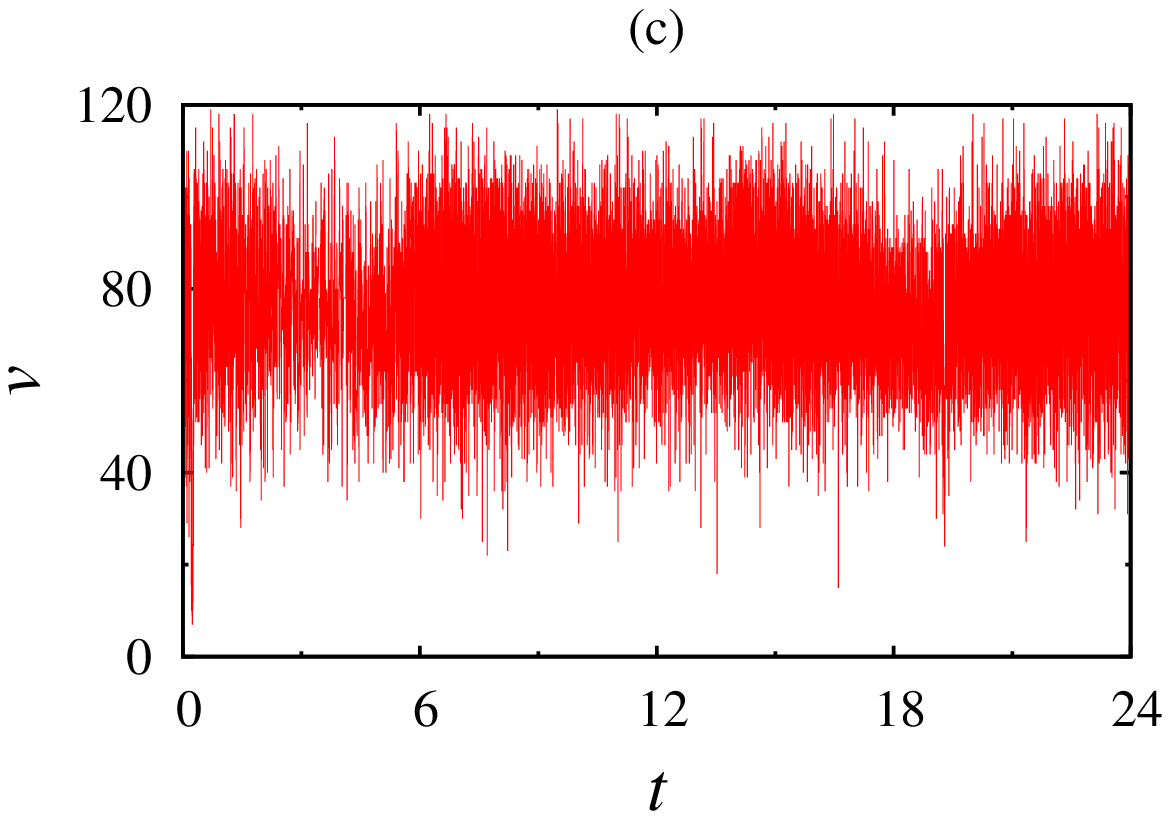}
  \includegraphics[width=0.71\columnwidth]{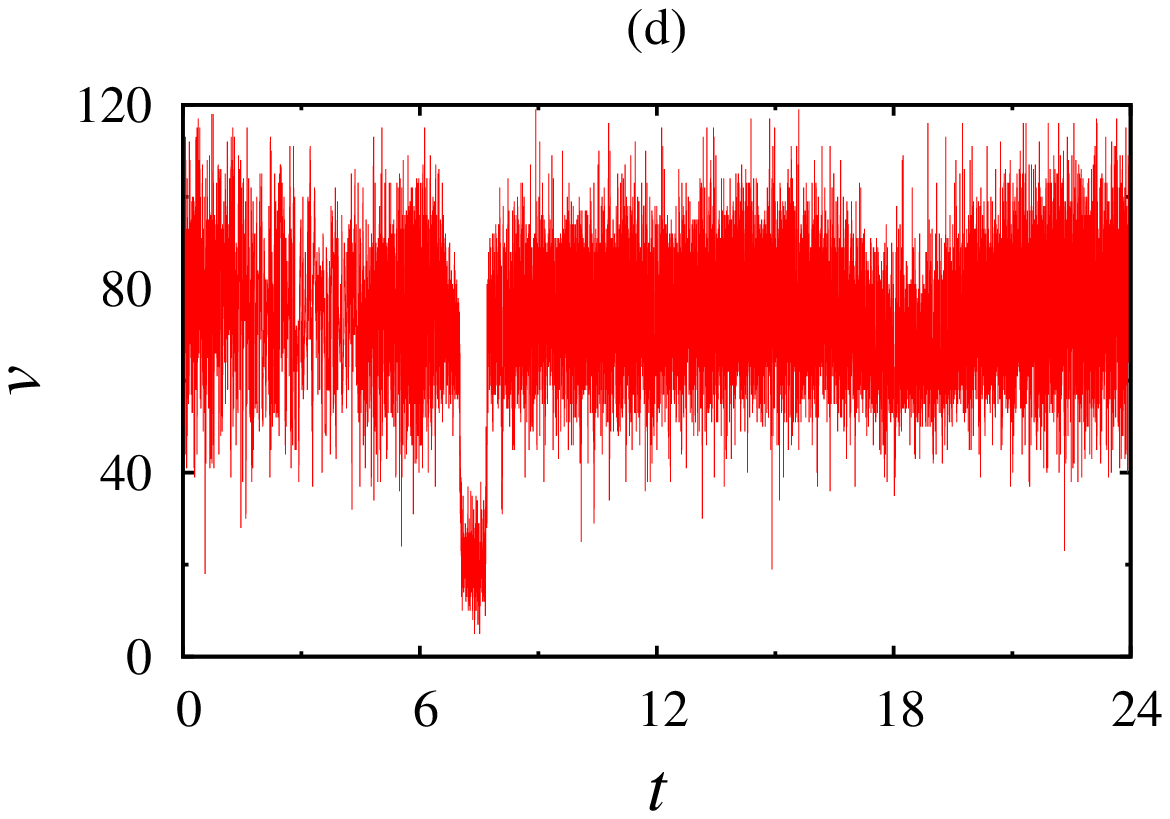}
  \includegraphics[width=0.71\columnwidth]{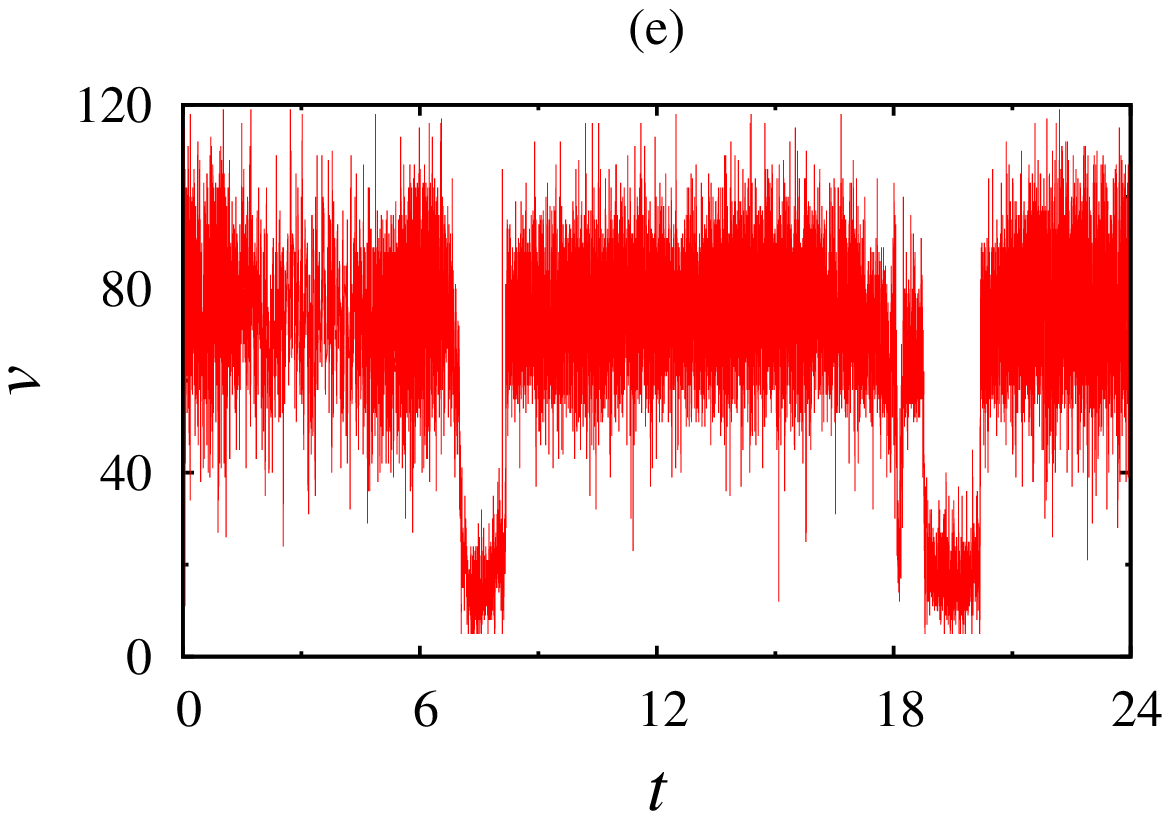}
  \caption{(Color online) Vehicle speed data at a cross section of Sayyad highway in Isfahan for the days (a) $8-10-2014$, (b) $9-10-2014$ (weekend),  (c) $10-10-2015$ (weekend), (d) $11-10-2015$ and (e) $12-10-2015$.    }
  \label{10}
\end{figure}

\begin{figure}[]
\begin{subfigure}
  \centering
  \includegraphics[width=\columnwidth]{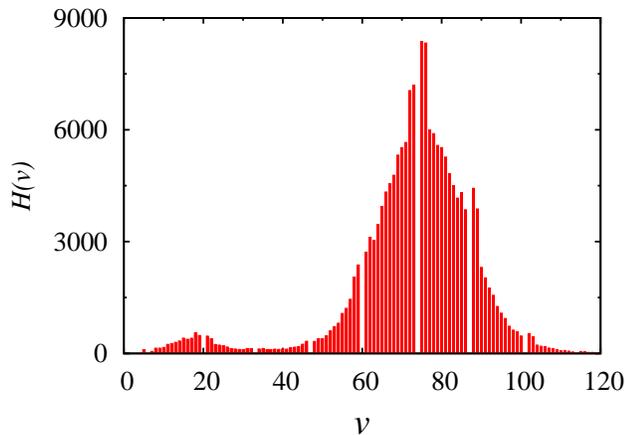}
  \caption{(Color online) Speed histogram for $5$ days of recorded data. }
  \label{Hv}
\end{subfigure}
\end{figure}

Figures \ref{10}-a,b,c,d,e demonstrates the speed time series during these $5$ days.
  Figure~\ref{Hv} shows the histogram of cumulative speed time series for all the $5$ days. It can be seen from this figure that there is a large peak in the speed histogram at $v\sim 75$ km/h, corresponding the free flow, and also a small peak corresponding  to the wide moving jam at $v\sim 20$ km/h.  

\begin{figure}
  \centering
  \includegraphics[width=\columnwidth]{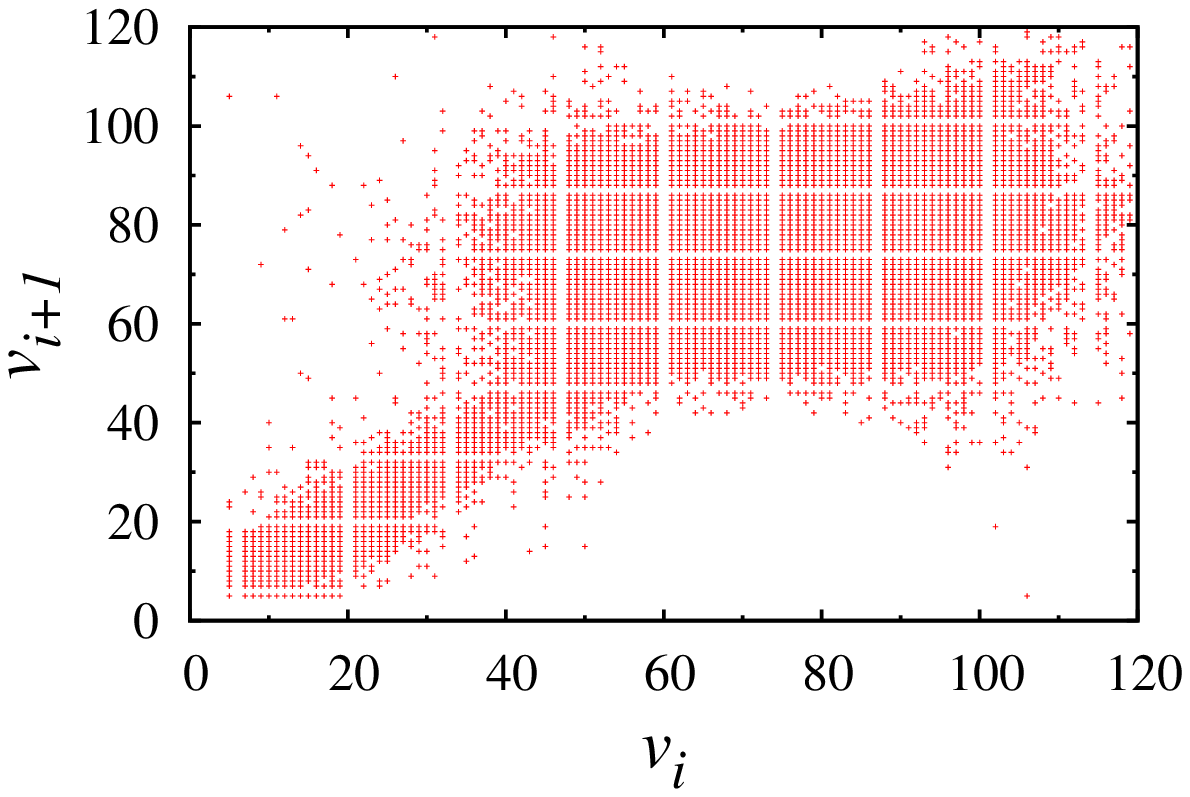}
 \includegraphics[width=\columnwidth]{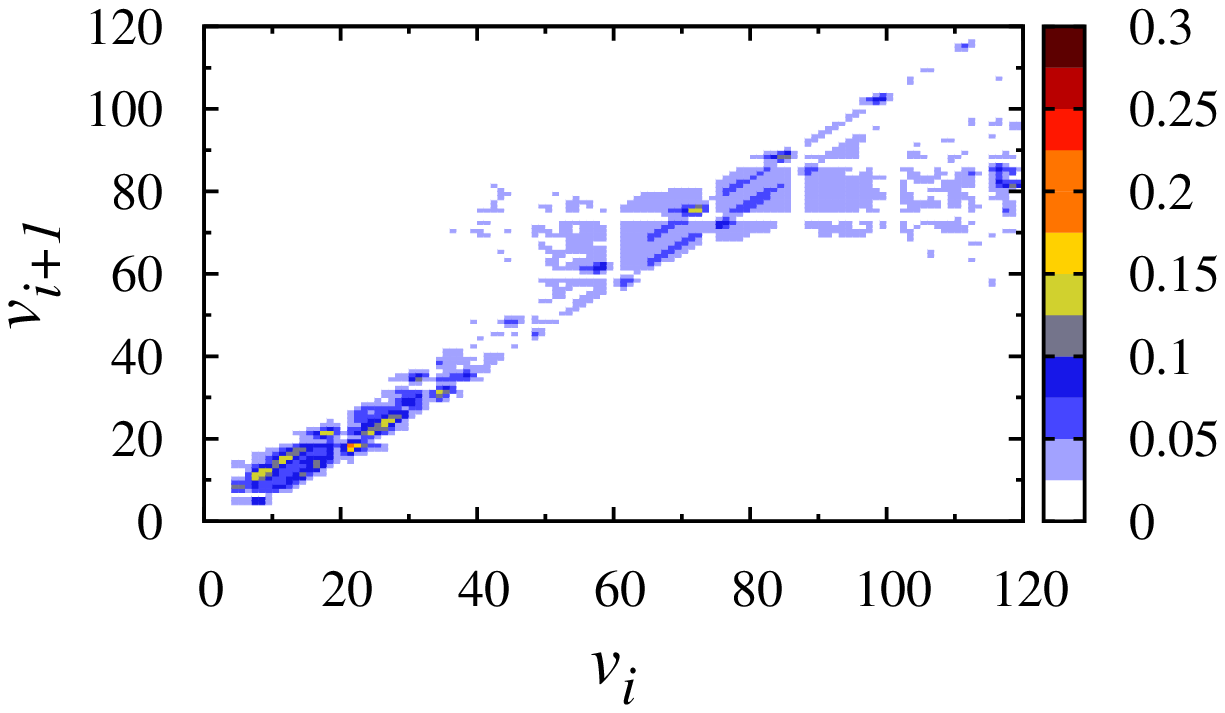}
  \caption{(Color online) ({\bf Top}): Scatter plot of speed at successive times,  $v_{i+1}$ versus $v_i$. ({\bf Bottom}): The density plot of the conditional probability $p(v_{2},i+1|v_{1},i)$.}
  \label{pv2v1}
\end{figure}

Figure \ref{pv2v1}  exhibits  the scatter plot of the speed at any two successive times $t+1$ versus $t $ (top panel) and also the conditional probability of finding speed at the value $v_2$ at time $t+1$ provided that its value at time $t$ be $v_1$ (bottom panel).

\begin{figure*}
\centering
  \includegraphics[width=\columnwidth]{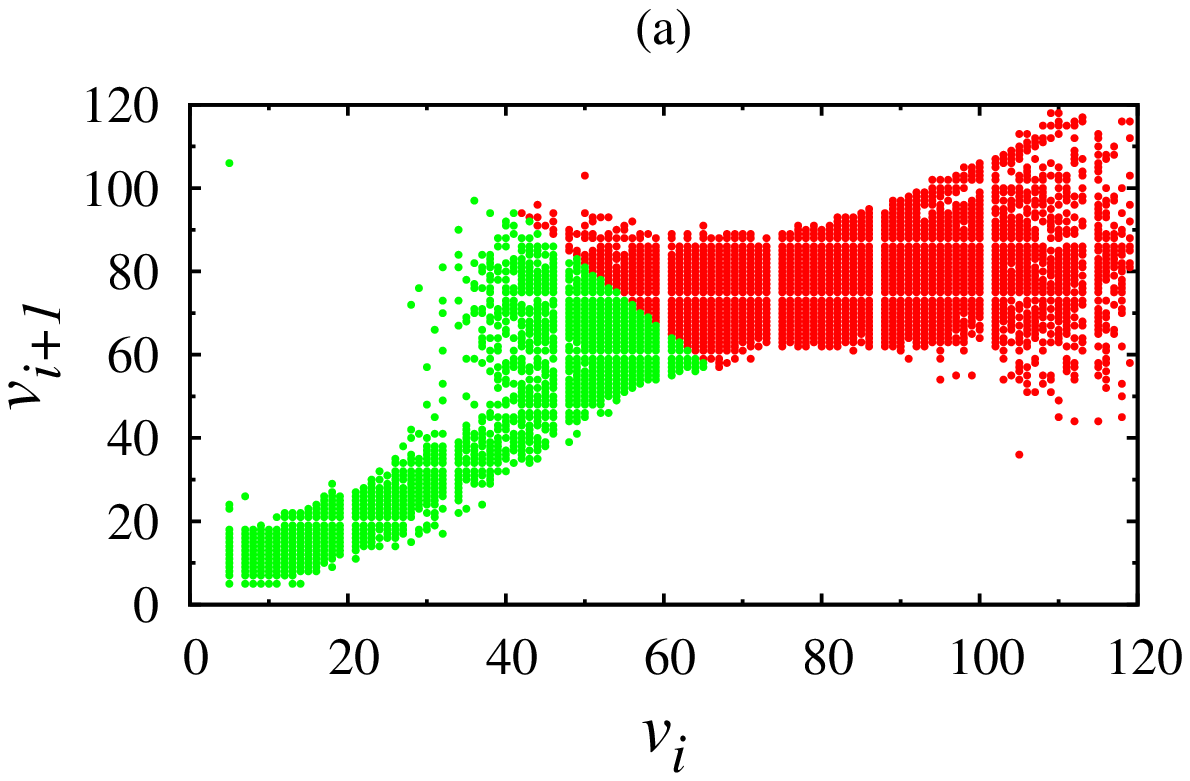}
  \includegraphics[width=\columnwidth]{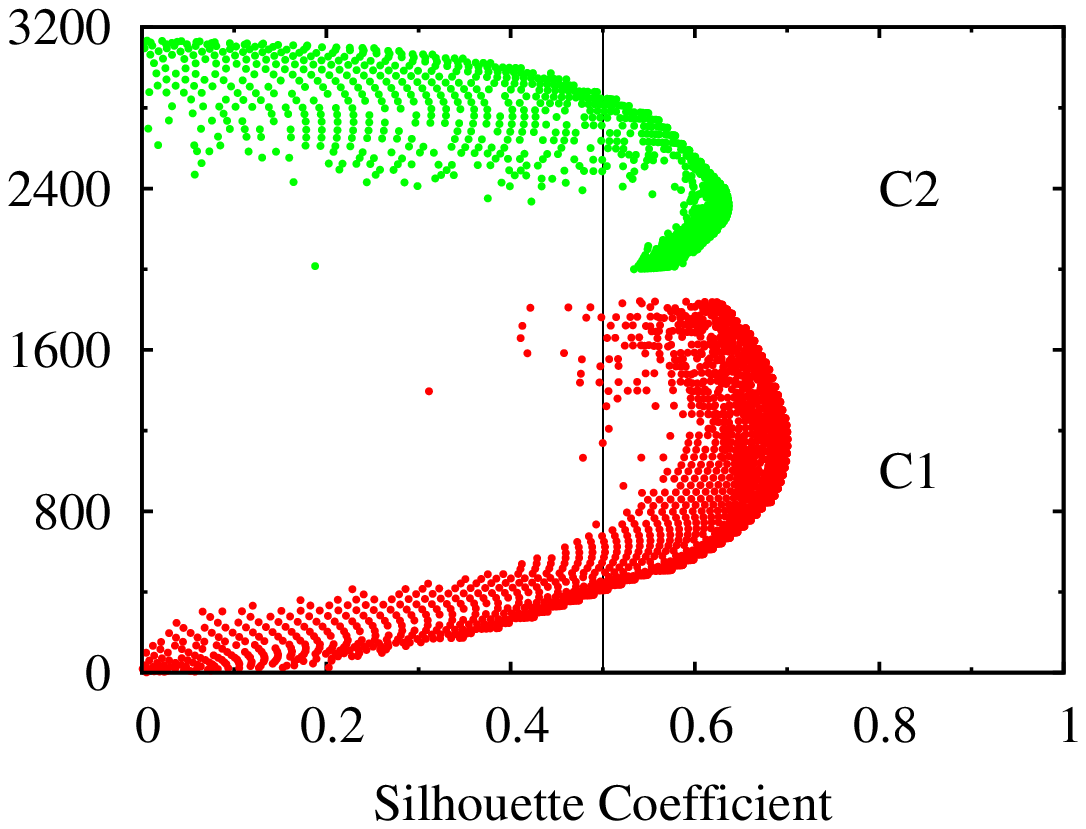}
  \includegraphics[width=\columnwidth]{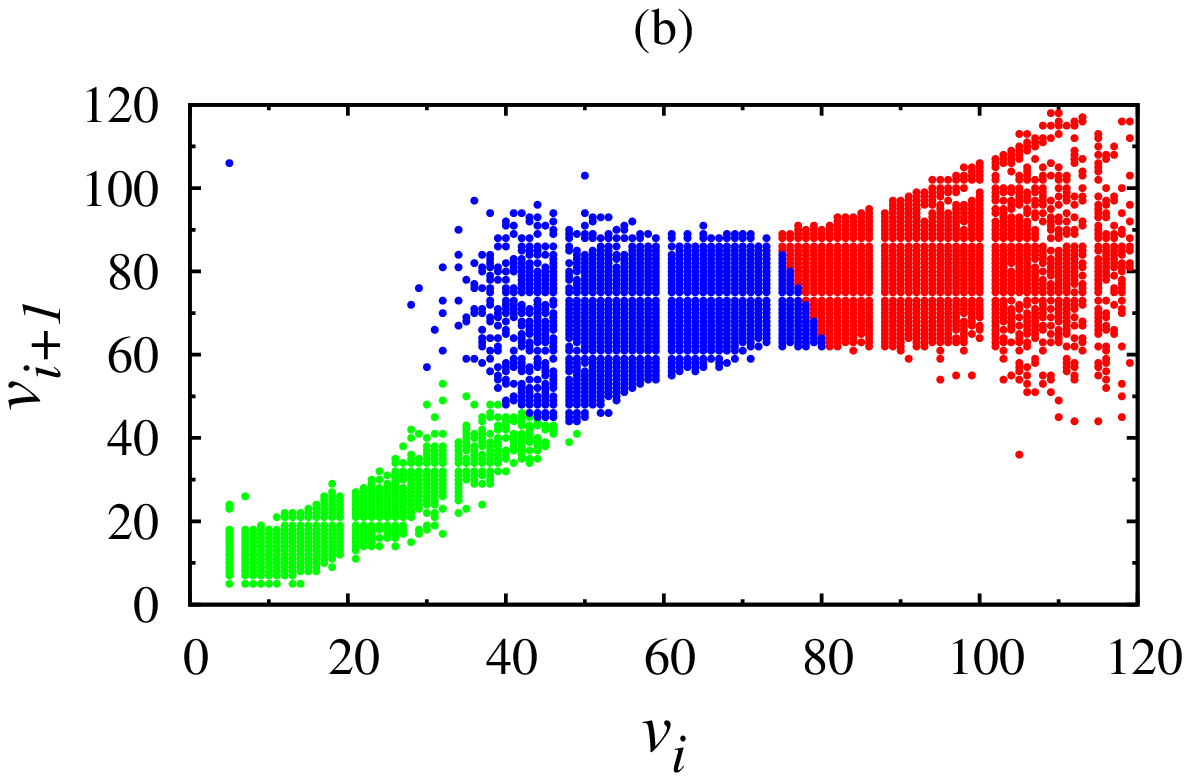}
  \includegraphics[width=\columnwidth]{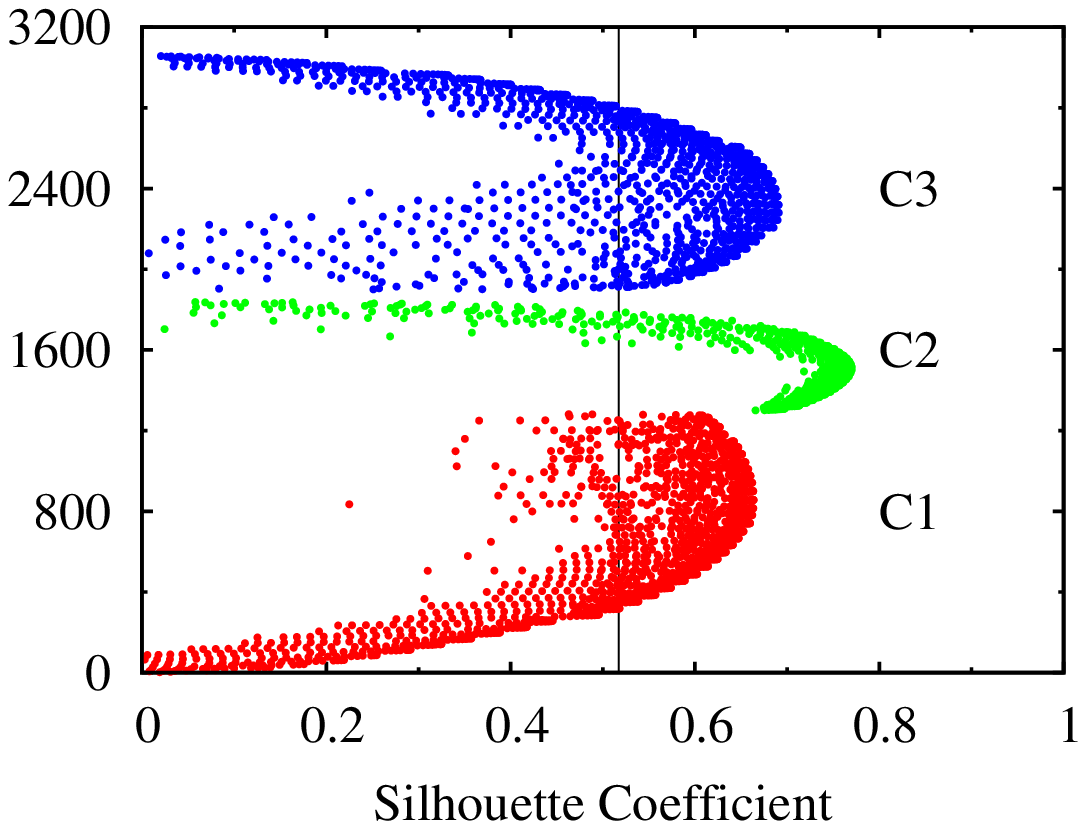}
  \includegraphics[width=\columnwidth]{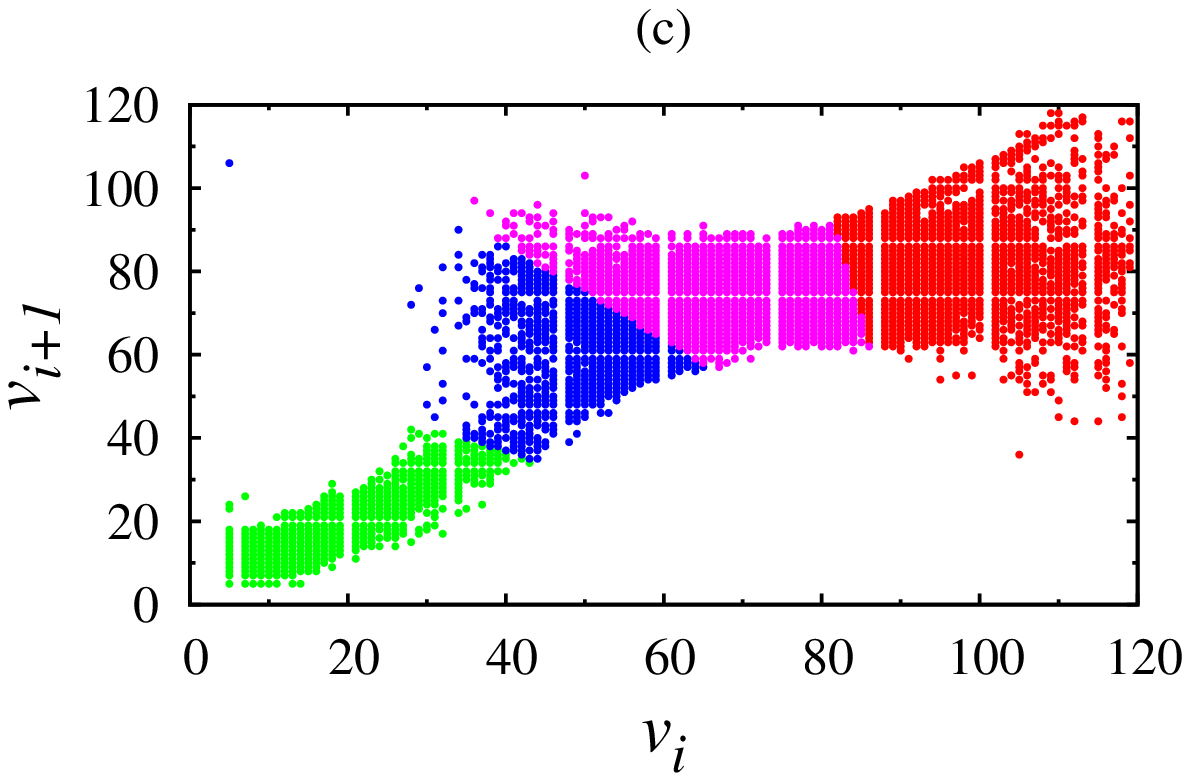}
  \includegraphics[width=\columnwidth]{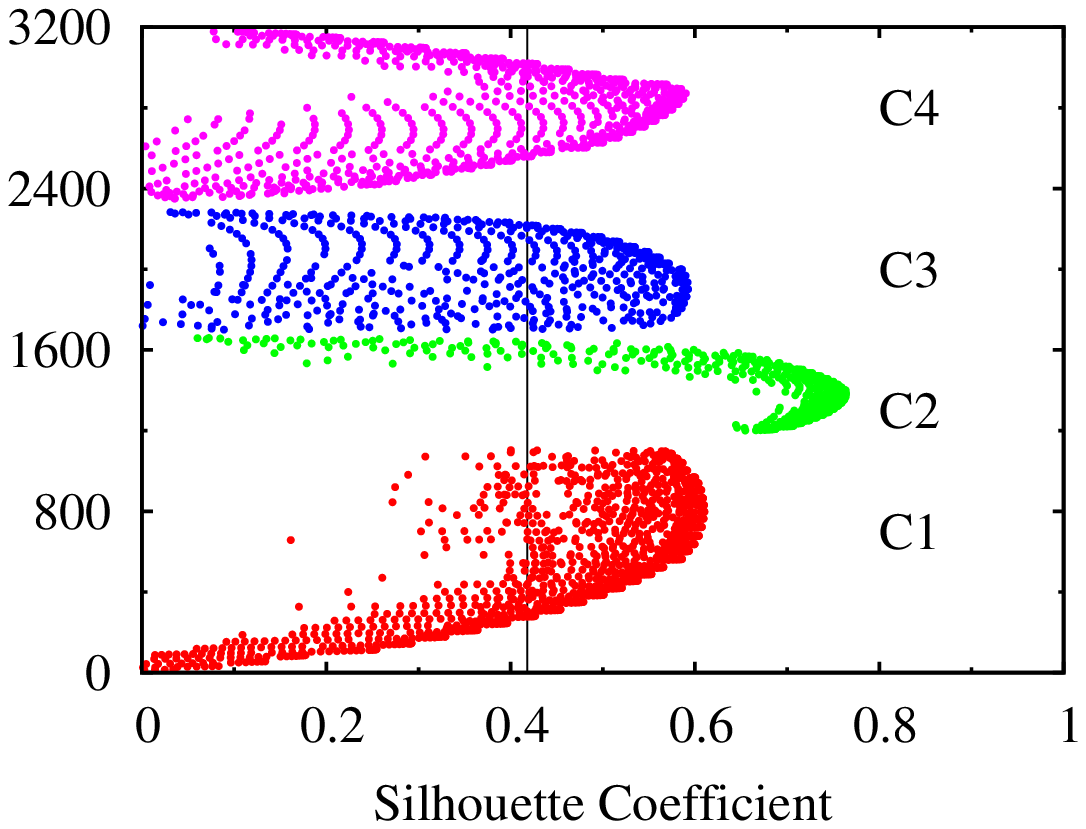}
\caption{(Color online) ({\bf Left}) $k$-means  clustering of data into (a) $2$, (b) $3$ and (c) $4$ clusters. ({\bf Right}) Silhouette coefficient for (a) $2$, (b) $3$ and (c) $4$ clusterings.}
  \label{$k$-means  clustering}
\end{figure*}


\begin{figure}
 \centering
  \includegraphics[width=\columnwidth]{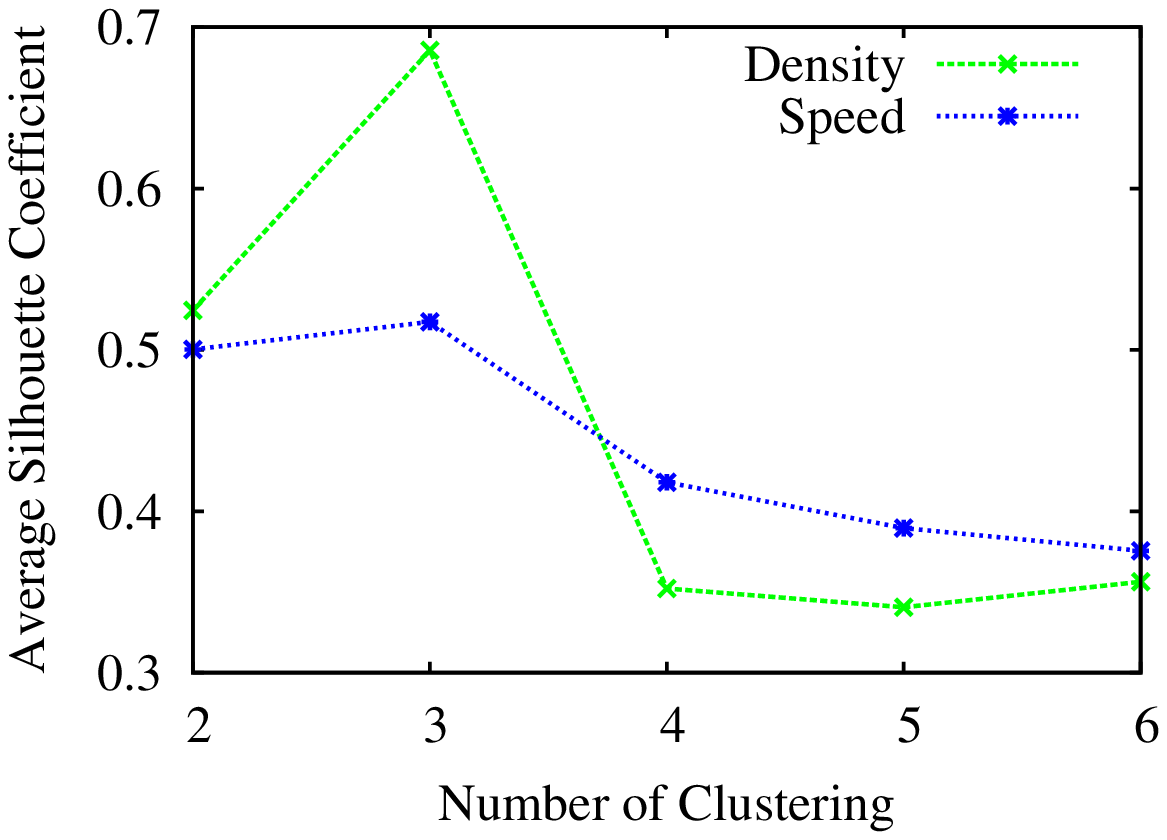}
\caption{(Color online) The average of Silhouette coefficients for different clustering numbers for scatter plot of speed (bluet) and density (green) time seires. }
\label{silave}
\end{figure}


\section{clustering of vehicle speed time series}
\label{k-mean}

In order to use the $k$-means  method for data clustering, first we assume a specific number of clusters and obtain the  corresponding Silhouette coefficient, then to obtain the natural number of clusters, we find the clustering for which the Silhouette coefficient is maximum.
In this work, we apply $k$-means algorithm to the scatter plot of speed data ($v_{i+1}$ versus $v_i$), shown in figure \ref{pv2v1}. 
To reduce the highly rare events which may results in unreal clustering, we first eliminate the data of the scatter  plot with low conditional probability, say $p(v2,i+1|v1,i) < 0.001$.   
 Results of clustering data to $2$, $3$ and $4$ groups are shown in the right panel of figures \ref{$k$-means  clustering}-(a), (b) and (c), respectively.   The Silhouette coefficients calculated for  data in each clustering  are also illustrated in the left panel of figures  \ref{$k$-means  clustering}-(a), (b) and (c). The vertical lines in silhouette  plots, show the average silhouette coefficient for that clustering.

The average Silhouette coefficient corresponding to different number of clustering are represented in figure \ref{silave} (blue symbols), showing this average reaches to a  maximum at $N=3$, whose message is  that the natural number of clustering  is $3$. 

\begin{figure}
  \centering
  \includegraphics[width=\columnwidth]{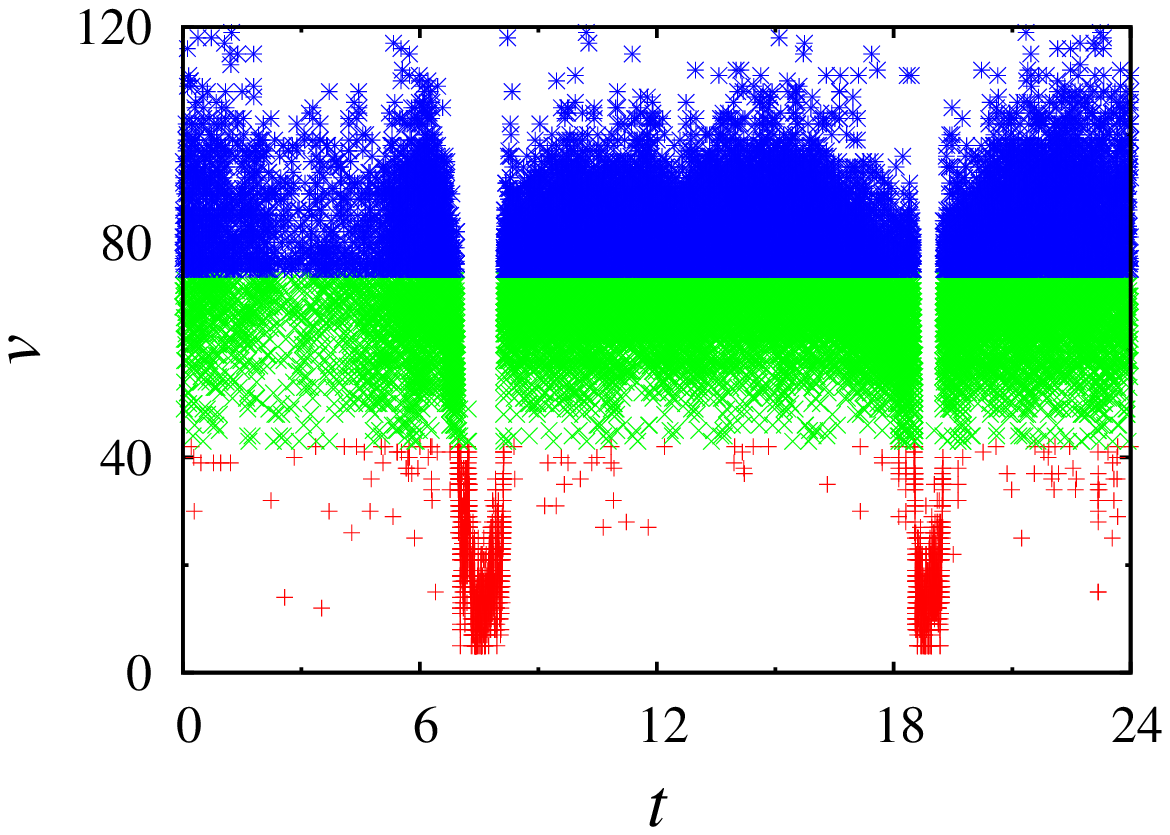}
  \caption{(Color online) $k$-means  clustering  of one day speed time series into $3$ clusters.   }
  \label{v-kmean}
\end{figure}

We also apply $k$-means  method for clustering of  the speed time series itself into three groups. As the results of such a clustering (shown in figure~\ref{v-kmean} for the first day) we find three groups: (i) $0 \leq v \lesssim 42$ km/h for wide moving jam,  (ii) $43$ km/h $\lesssim v \lesssim 74$ km/h for the synchronized phase and (iii) $74$ km/h $\lesssim v \lesssim 120$ km/h for the free flow.  

\begin{figure}
  \centering
  \includegraphics[width=\columnwidth]{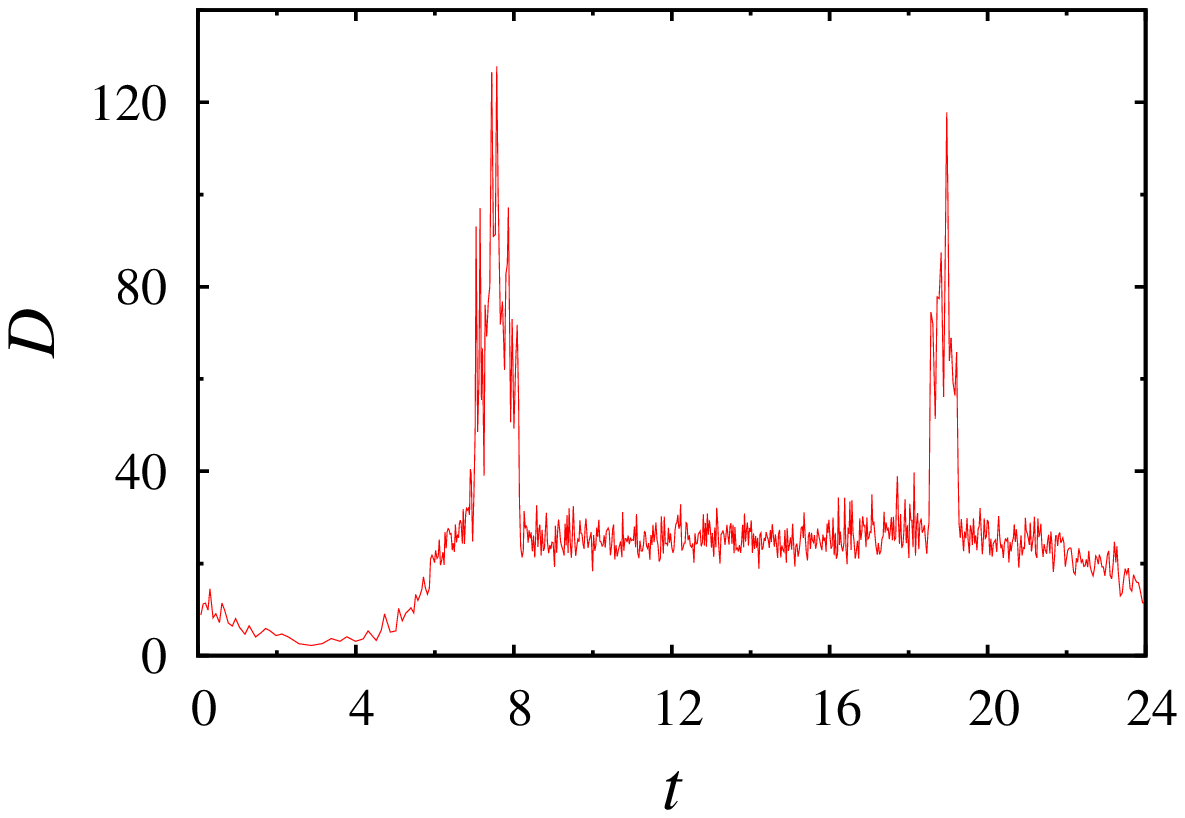}
  \caption{(Color online) Vehicle density time series for the day 9-10-2014 corresponding to the speed time series shown in
  Fig~\ref{10}-(a).   }
  \label{D-series}
\end{figure}

It is also possible to obtain a rough estimation of density  from the speed time series. 
Consider a time window in the speed series of length $N$ vehicles, starting from the time $t_i$. Then the density at the time $t_i$ ($D_i$) can be estimated as
\begin{equation}
D_i=\frac{N}{\Delta t_{i} \bar{V}_{i}},
\label{den-eq}
\end{equation}
in which $\Delta t_{i}$ denotes the time duration of passing $N$ vehicles starting from the time $t_i$, and
 ${\bar V}_{i}$ is the average speed of the vehicles in this window. Here we choose the window of $N=50$. 
 Figure~\ref{D-series} represents the density times series for one day of the  highway traffic flow, in which the wide moving jams
 appear as sharp peaks. 
 
\begin{figure}
  \centering
  \includegraphics[width=\columnwidth]{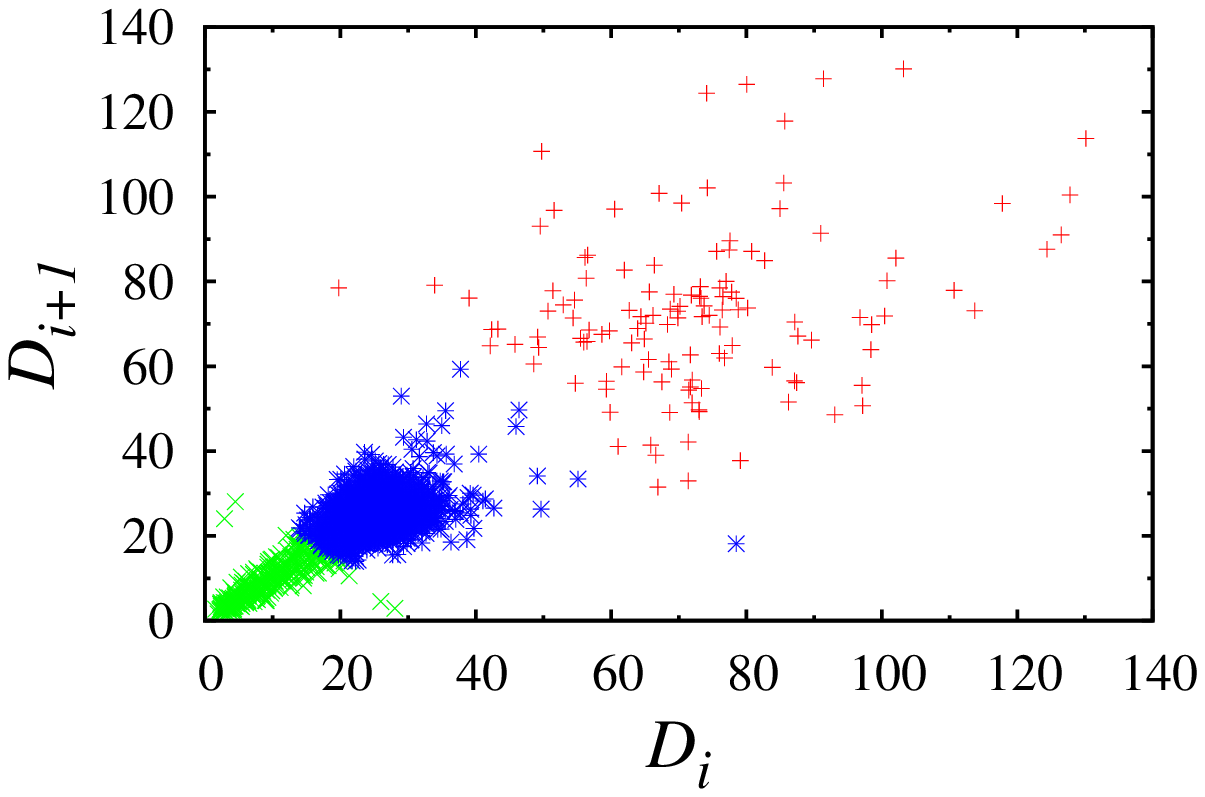}
  \caption{(Color online) $k$-means  clustering clustering for $k=3$ on density scatter plot.   }
  \label{den3-kmean}
\end{figure}

\begin{figure}
  \centering
  \includegraphics[width=\columnwidth]{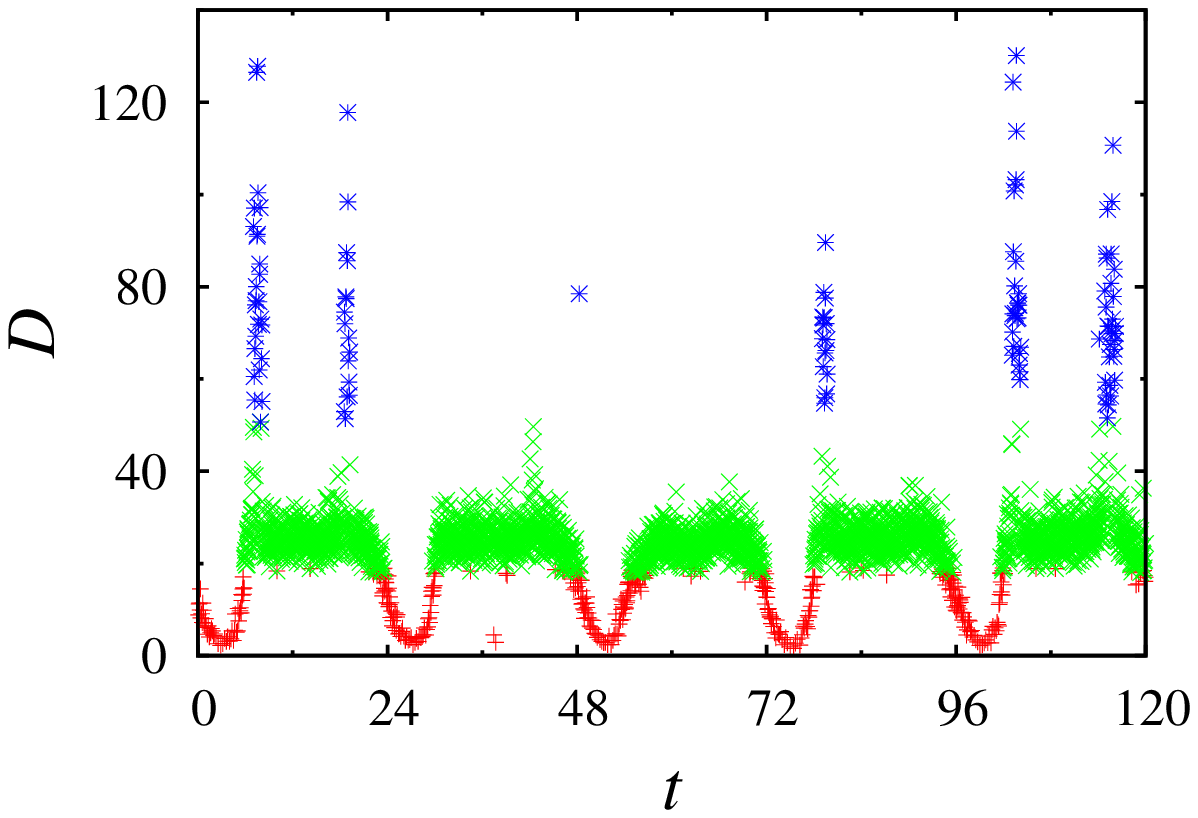}
  \caption{(Color online) $k$-means  clustering clustering for k=3 on density series.   }
  \label{den3-time}
\end{figure}

Figure~\ref{den3-kmean} shows the results of the implementation of $k$-means clustering method on the density scatter plot for successive time windows ($D_{i+1}$ vs $D_{i}$) into three clusters.  The silhouette analysis, (figure \ref{silave}, green symbols), again reveals $3$ natural clustering for the uninterrupted traffic flow. As it makes us sure that three clustering is the best natural number of clusters for density, we can find the values of the densities  separating those  three regions  by simply applying the three clustering $k$-means  on the density time series. The results are exhibited  in figure~\ref{den3-time}  and show that the three groups of density can be classified as  (i) $0\leq D\lesssim18$/km for free flow,  (ii) $18$/km$\lesssim D \lesssim 50$/km for the synchronized phase and (iii) $D > 50$/km for the wide moving jam. 


\begin{figure}
  \centering
  \includegraphics[width=\columnwidth]{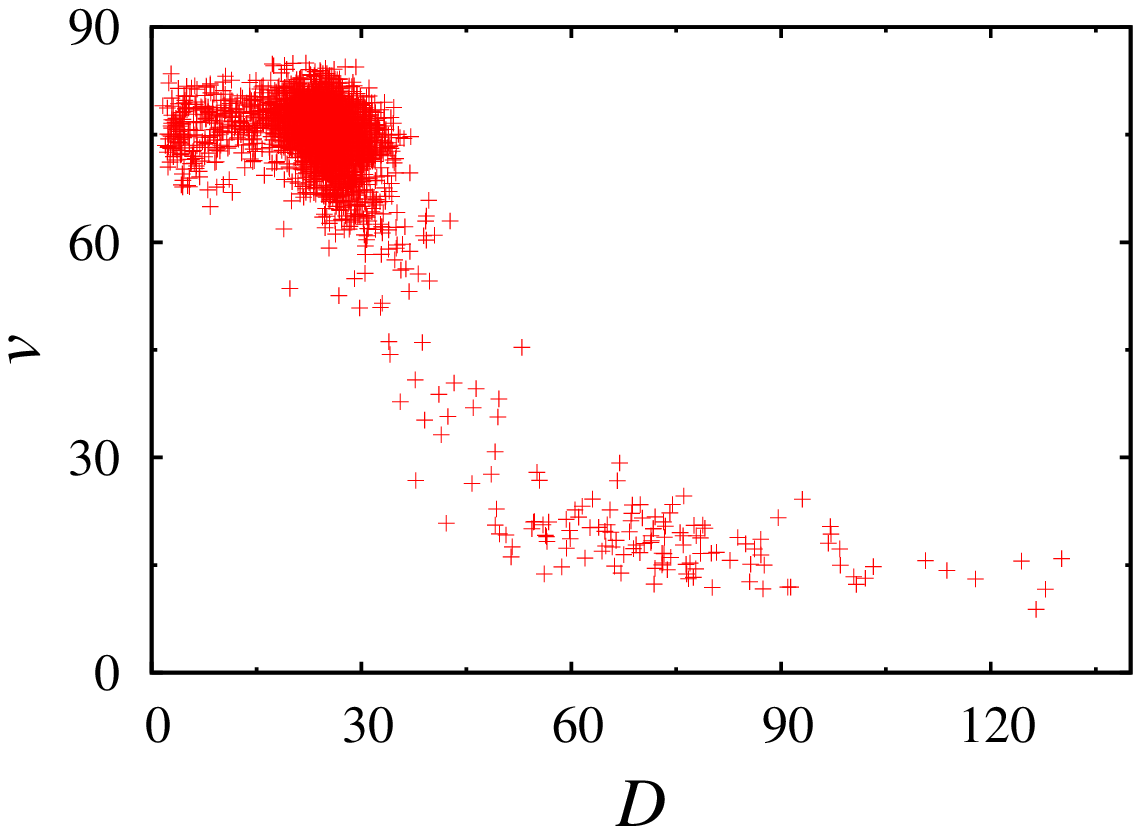}
  \caption{ (Color online) Fundamental diagram of speed versus density. }
  \label{dv}
\end{figure}

To wrap up this section, we discuss the fundamental diagram of speed versus density illustrated in figure~\ref{dv}. The three phases of the uninterrupted traffic flow are also apparent in this figure. It can be seen that the free flow and wide moving jam can be identified by the region in the fundamental diagram ($v-D$), where speed is roughly a constant value independent of  density. In this case, for the density less than $\sim 18$/km (free flow) the average speed of the vehicles is about $75$ km/h, while for vehicle density larger than $\sim 50$/km (wide moving jam) the average speed is roughly $20$ km/h. Between these two limits (synchronized) the average speed decreases almost linearly with increasing density. 

\section{Shannon Entropy analysis of speed time series} 
\label{Shannon}

In this section, we employ the Shannon entropy analysis to extract information contained in the time series. 
Thanks to $k$-means clustering method which gives us the boarder values of speed separating the three traffic phases,  the 
information of the time series coming from each state can be calculated as 

\begin{eqnarray}
&&H_{J}=-\Sigma_{v=0}^{v_j}p (v) \log p(v) \nonumber \\
&&H_{S}=-\Sigma_{v=v_j}^{v_s}p(v) \log p(v) \nonumber \\
&&H_{F}=-\Sigma_{v=v_s}^{v_m}p(v) \log p(v), 
\end{eqnarray}
where $p(v)$ is the probability of finding speed in the interval $[v, v+1 ]$ and $H_J$, $H_S$ and $H_F$ denote the Shannon entropy corresponding to wide moving jam, synchronized and free flow, respectively. $v_s$ is the transition speed separating  wide moving jam from synchronized, $v_j$ is the speed at the border of synchronized and free flow states and $v_m=120$ km/h is maximum recored  speed. Based on $k$-means clustering analysis we set  $v_j=42$ km/h and $v_s=74$ km/h.  

\begin{figure}
 \centering
 \includegraphics[width=\columnwidth]{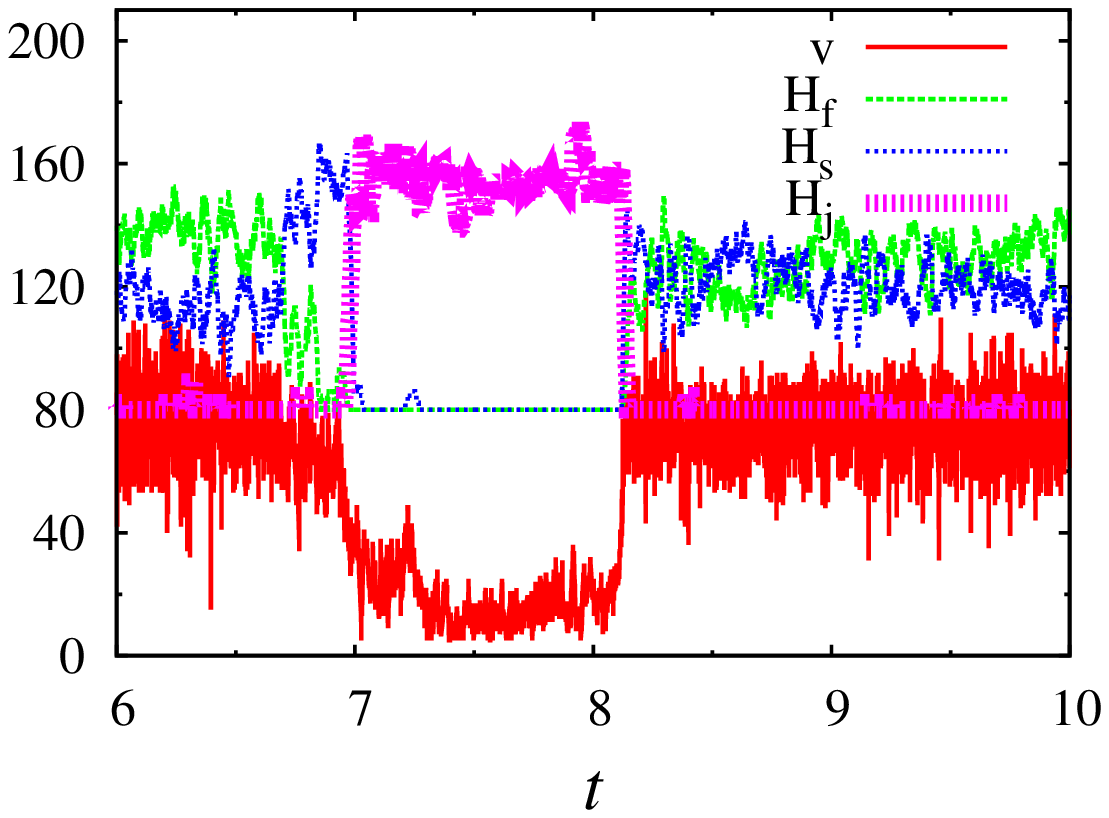}
   \caption{(Color online) Shannon entropies corresponding to wide moving jam ($H_J$), synchronized ($H_S$) and free flow ($H_F$) states of  speed time series around a wide moving jam. For better visualization, the entropies are multiplied by the factor $3$. }
 \label{shannon}
\end{figure}

We are interested in determining the instant state of traffic flow, hence we first divide the time series into the finite time windows of the order of correlation length of time series which is typically $50$. In figure~\ref{shannon} the three entropies are plotted close to the occurrence  of a  wide moving jam. This figure clearly shows the transition between different phases of traffic. As can be seen in  figure~\ref{shannon}, long before the jam the dominant entropy is $H_F$, an indication of free flow phase, however  before the onset of jam, synchronized  entropy $H_S$ rises and  become dominant over the other two and finally at the onset of jamming it is the wide moving jam entropy $H_j$ which acquires  the largest value.  Right after the end of jamming state, however, none of the synchronized or free flow entropies are dominant over each other, meaning that during time periods right after the end of the jamming phase, there could be  coexistence between F and S  phases.

 \section{conclusion}
\label{conclusion} 

In summary, we found that $k$-means clustering  method is a powerful instrument to characterize different states of an uninterrupted  traffic flow. Applying this method to both the speed times series  and also the density time series calculated form speed data, we sound that the three phases  wide moving jam, synchronized and  free flow come up as the natural clusters. This method gives us the speed and densities at which the transitions between these states occur. Implementation of  Shannon entropy analysis over finite time windows,  enabled us to find the short time states of traffic flow. From the Shannon entropy analysis we find that in going from free flow to a wide moving jam, first a transition occurs  between the F and S and then from S to J. Nevertheless, right after the end of a  wide moving jam, synchronized and free flow states could coexists.  

Our findings in this work are consistent with three-phase theory and we hope that may open avenues through optimization for highwy traffic programming.

\end{document}